\documentclass[a4paper]{mem}
\usepackage{natbib}
\usepackage{graphicx}
\begin{document}
   \title{Obscured AGN in Multiwavelength Surveys
}

   \author{Ezequiel Treister \inst{1,2}, 
   C.M. Urry \inst{1}, Jeff Van Duyne \inst{1}\\ 
          \and
          the GOODS AGN Team
}

   \offprints{E. Treister}
\mail{P.O. Box 208101 New Haven, CT 06520-8101 USA}

   \institute{Yale Center for Astronomy \& Astrophysics, Yale University, New Haven, CT
\email{treister@astro.yale.edu}\\ 
              \and  Departamento de Astronom\'{\i}a, Universidad de Chile, Santiago, Chile\\
             }

   \abstract{ Using a simple unification model for AGN,
   with a fixed ratio of
   obscured to unobscured AGN of 3:1, we explain the X-ray,
   optical and infrared properties of the X-ray sources in
   the GOODS fields. That is, the GOODS data are consistent
   with the existence of a large population of obscured AGN 
   out to high redshifts. About half of these are so obscured that
   they are missed even by hard X-ray observations, and
   are detected only in deep infrared observations. The
   previously reported decreasing trend of obscured to
   total AGN ratio with increasing X-ray luminosity can be explained 
   entirely as a selection effect.

   }
   \authorrunning{E. Treister et al.}
   \titlerunning{Obscured AGN in Multiwavelength Surveys}
   \maketitle
%

\section{Introduction}

The long-standing problem of explaining the nature of the
extragalactic X-ray background (XRB) is finally being
solved. Recently, the deepest Chandra and XMM observations
have resolved $\sim 70-90$\% of the XRB into point sources,
the vast majority of them identified as active galactic
nuclei (AGN). However, some problems remain.  The hard
spectrum of the XRB, much harder than that of the typical
unobscured AGN \citep{mushotzky00}, means that most of the
emission comes from \emph{obscured} AGN \citep{setti89}, yet
the observed ratio of obscured to unobscured AGN in the
XMM/Chandra deep fields is only $\sim2:1$, much lower than
the 4:1 ratio inferred from population synthesis models
\citep{gilli01}. Also, such models predict a redshift
distribution that peaks at $z\sim 1.4$, while the deep field
AGN distributions peak at a much lower redshift, $z\sim 0.7$
\citep{hasinger02}.

To find the large numbers of obscured AGN predicted by XRB
models has proved to be a very hard task. Wide-area optical
surveys like the Sloan Digital Sky Survey, which discovered a
large number of unobscured AGN, are not very efficient
for detecting obscured AGN given their low optical fluxes and lack
of broad emission lines. Therefore, multiwavelength surveys ---
in particular at hard X-rays and infrared wavelengths, where
most of the obscured AGN emission is found --- are required to
obtain a more complete census of the AGN population.

This was one of the main motivations for the Great
Observatories Origin Deep Survey (GOODS), which consists of
deep imaging in the far infrared with the Spitzer Space
Telescope \citep{dickinson02} and in the optical with the
Hubble Space Telescope \citep{giavalisco04} on the
footprints of the two deepest Chandra fields
(\citealp{giacconi01,brandt01}). The total area is roughly
60 times larger than the original Hubble Deep Field and
reaches nearly the same optical flux limit.
With extensive coverage over
five decades in energy from 24~$\mu$m to 8~keV, the GOODS
survey is well suited to find a high-redshift population of
obscured AGN if they exist. A complementary approach, given
the relatively low surface density of AGN (compared to
normal galaxies), is to target higher luminosity AGN over a
wider area of the sky, an approach followed by the
CYDER \citep{castander03} survey, for example.  Here we describe the
multiwavelength properties of AGN detected in X-rays in the
GOODS North and South fields, which represent an order of
magnitude more objects than in most previous works.

\section{The Model}

Much work on black hole demographics begins with the AGN
found in a given survey, correcting where possible for
selection biases to infer the underlying population.  If
selection effects are strong, however, one ends up making
large extrapolations using little information. We therefore
took a different approach: we asked, if there is a
substantial population of obscured AGN, what would be seen
in a deep multiwavelength survey like GOODS? Our work is
presented by \citet{treister04a}; here we briefly describe
our assumptions and results.

To derive the number counts at any wavelength we start with
a hard X-ray luminosity function, an assumed cosmic
evolution, and a library of spectral energy
distributions. We use the hard X-rays as a starting point
because observations from 2-10 keV in the rest frame are
less affected by obscuration and therefore provide a less
biased view of the AGN population. In this work, we use the
luminosity function and evolution of Ueda et al. (2003,
hereafter U03). We construct SEDs as a function of only two
parameters, the intrinsic hard X-ray luminosity and the
neutral hydrogen column density ($N_H$) along the line of
sight. For the X-ray spectrum, a simple power law with slope
$\Gamma=1.9$ plus photoelectric absorption with solar
abundances was assumed. In the optical, we use the SDSS
composite quasar SED \citep{vandenberk01}, absorbed using
Milky-Way type extinction plus an $L_*$ elliptical host
galaxy. In the infrared region, the dusty torus
emission models of \citet{nenkova02} were used.

The dependence of the luminosity function on the column
density is calculated separately using an ``$N_H$ function''
presented in Equation 6 of U03, which is based on the
relative number of sources at each $N_H$ observed in their
sample. The $N_H$ distribution was calculated based on the
unified paradigm, in which the torus has a fixed geometry
and dust distribution. Our model comprises this $N_H$
function, combined with U03 luminosity function and
evolution and the previously described library of AGN SEDs.

\section{Results}

   \begin{figure*}
   \centering
   {\includegraphics[clip=true,height=6.3cm]{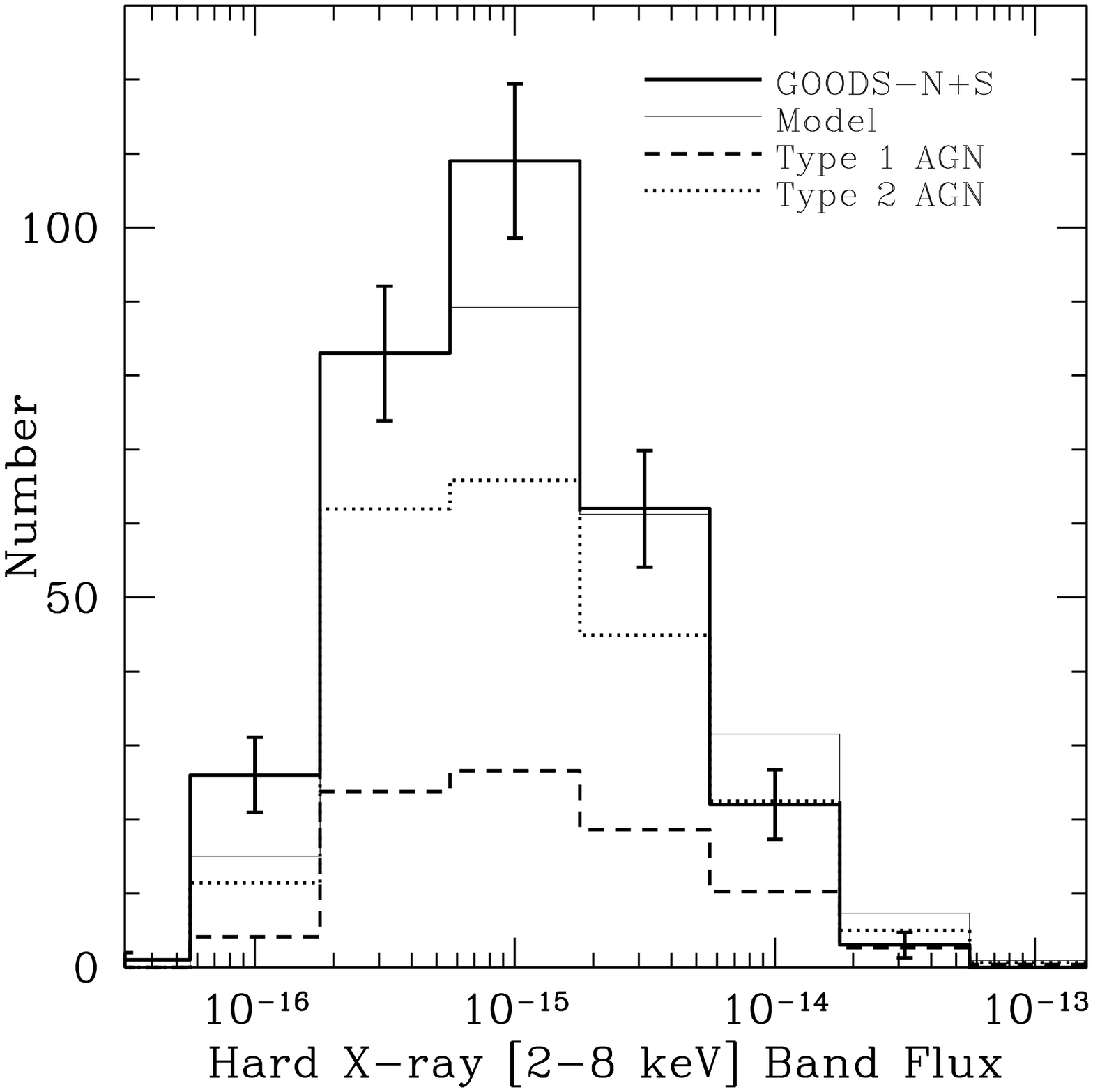}
   \includegraphics[clip=true,height=6.3cm]{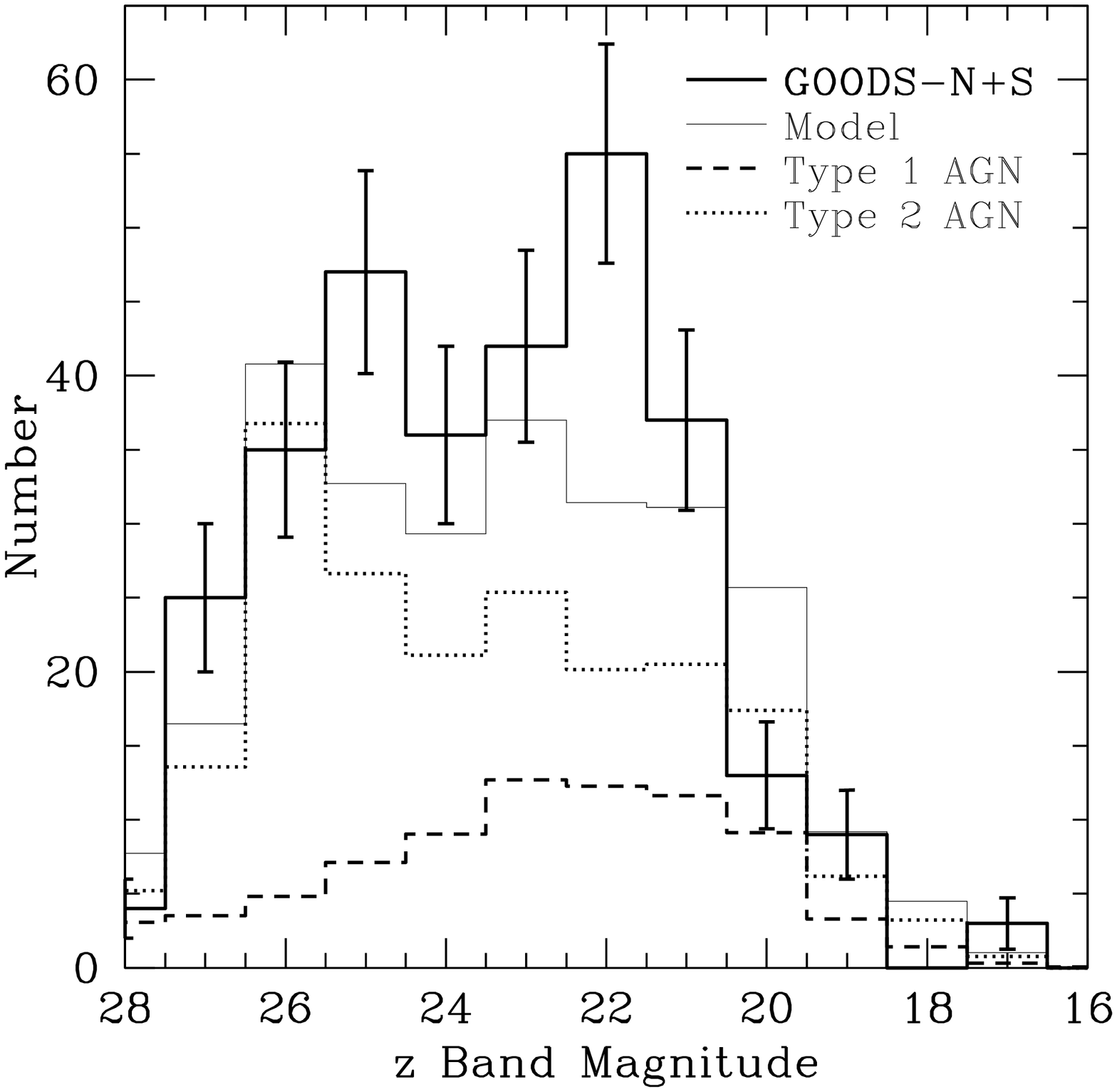}}
\caption{X-ray
   fluxes and z-band magnitudes of GOODS AGN.  {\it (Left)}
   Hard X-ray flux (2-8.0~keV) distribution for X-ray
   sources in the combined GOODS fields
   ({\it heavy solid line}, compared to number counts
   calculated from a simple unification model ({\it light
   solid line}). Individual contributions from unobscured
   ({\it dashed line}) and obscured ({\it dotted line}) AGN
   are shown separately.  {\it (Right)} Distribution of observed $z$-band magnitudes for
   GOODS X-ray sources ({\it heavy solid
   line}), compared to model distribution ({\it light solid
   line}).} 
   \label{fig1} 
\end{figure*}

Using the model described in the past section, we computed
the expected hard X-ray flux and optical magnitude
distribution for the X-ray sources in the GOODS
fields. Compared to the observed distribution, the agreement
is remarkable, showing that this model is able to account
for the observed AGN population in these deep Chandra
fields, as shown in Fig. 1. Also, the predicted and observed
redshift distributions are consistent once that selection
effects in the observed sample are accounted for. The need
for optical spectroscopy generates the largest selection
effect in this sample, since most obscured AGN are faint in
the optical, meaning it is much harder to obtain
spectroscopic redshifts for them. 

The GOODS fields will be observed by Spitzer in the four
IRAC bands (3.6, 4.5, 5.8 and 8 microns) and in the MIPS 24
microns band. Currently, we have obtained data in the North
and South fields in the IRAC bands and only in the North
field with MIPS. Preliminary results show that both the
overall shape and normalization of the predicted and
observed distributions agree very well once that the
predicted infrared fluxes are reduced by a constant factor
of $\sim 2$. This may indicate that the overall torus
emission was overestimated, suggesting a smaller mass for
the average AGN dust torus. Also, this model suggests that
only half the AGN in the GOODS fields are 
detected in the deep Chandra X-ray observations, with the
remaining half being very obscured, moderate luminosity
AGN. All these missed AGN will be luminous enough to be
easily detected in the GOODS Spitzer observations.

Previous work (\citealp{hasinger03};U03) found an
apparent decrease of the ratio of obscured to total AGN with
increasing X-ray luminosity. Again, this can be
explained as a selection effect against the determination of
spectroscopic redshifts for obscured AGN at higher
luminosities, in particular at higher redshifts were most of
them are found (Fig.~2; \citealp{treister04b}).

\begin{figure*} 
\centering
\includegraphics[angle=-90,scale=0.325]{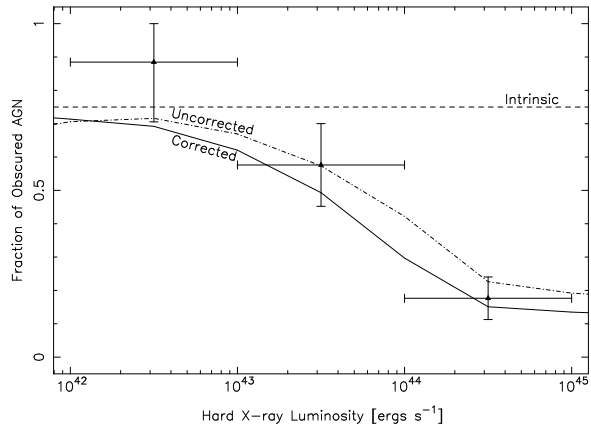}
\caption{The fraction of obscured AGN 
versus X-ray luminosity, shown here for X-ray sources in
the combined CYDER and GOODS surveys, appears
to decrease with X-ray luminosity, as has been
seen in previous work (\citealp{hasinger03};U03).
({\it Dot-dashed line:}) Predicted
trend for a unified model with a constant intrinsic ratio
of obscured to total AGN of 3:4 ({\it dashed
line}), considering only objects with optical magnitude $R<24$ mag
(i.e., the optical cut for spectroscopy) and
ignoring the (observationally unknown)
effects of obscuration and K correction 
in calculating the X-ray luminosity. {\it Solid
line:} Predicted trend after correcting the intrinsic hard
X-ray luminosity for obscuration and redshift effects.}
\label{fig2}
\end{figure*}

\section{Conclusions}

We compared the
multiwavelength properties of the hard X-ray sources
detected in the GOODS fields with a simple AGN unification
model and showed that the data are consistent with a large
number of obscured AGN at high redshifts. Optical
spectroscopy is very hard for such AGN,
even with modern 8-m class
telescopes, and thus they are normally excluded from AGN surveys
that rely on optical spectroscopy to determine redshifts
(and thus luminosities). Once this selection effect
is considered, the predicted and observed optical and hard
X-ray flux distributions and redshift distributions are in
agreement.

According to our predictions and early observations all the
AGN detected in X-rays will be bright in the infrared bands
and will be detected in the deep Spitzer observations of the
GOODS fields. However, $\sim 50$\% of the AGN in the field
were not detected in these deep X-ray observations, all them
being obscured AGN at high redshifts; these will be
detected only in infrared observations, where most of the
absorbed energy is re-emited.

We also found that the previously reported decrease in the
obscured to total AGN ratio with increasing X-ray luminosity
can be explained as a selection effect against the
detection of obscured AGN at high redshift, where most of
the high luminosity AGN can be found.

\begin{acknowledgements}
ET thanks the LOC of the conference for providing
significant financial support. This work was supported in
part by NASA grant HST-GO-09425.13-A and by Fundaci\'on
Andes.
\end{acknowledgements}

\bibliographystyle{aa}

\end{document}